\documentclass[conference]{IEEEtran}
\IEEEoverridecommandlockouts
\usepackage{cite}
\usepackage{amsmath,amssymb,amsfonts}
\usepackage{algorithmic}
\usepackage{graphicx}
\usepackage{tabularx}
\usepackage{textcomp}
\usepackage{xcolor}
\usepackage{dblfloatfix}
\usepackage[bookmarks=false]{hyperref}
\def\BibTeX{{\rm B\kern-.05em{\sc i\kern-.025em b}\kern-.08em
    T\kern-.1667em\lower.7ex\hbox{E}\kern-.125emX}}
\newcolumntype{Y}{>{\centering\arraybackslash}X}

\begin{document}

\title{Prediction of Energy Consumption for Variable Customer Portfolios Including Aleatoric Uncertainty Estimation
\thanks{This study was funded by the EU(Eurostars, \textit{ESB} project, grant: EUS-2019113348}
}

\author{\IEEEauthorblockN{Oliver Mey, André Schneider, Olaf Enge-Rosenblatt}
\IEEEauthorblockA{\textit{Fraunhofer IIS/EAS, Fraunhofer Institute for Integrated Circuits,} \\
\textit{Division Engineering of Adaptive Systems}\\
Dresden, Germany \\
e-mail: oliver.mey@eas.iis.fraunhofer.de}
\and
\IEEEauthorblockN{Yesnier Bravo, Pit Stenzel}
\IEEEauthorblockA{\textit{R\&D Department} \\
\textit{Bettergy S.L.}\\
Málaga, Spain \\
e-mail: ybravo@bettergy.es }
}

\maketitle

\begin{abstract}
Using hourly energy consumption data recorded by smart meters, retailers can estimate the day-ahead energy consumption of their customer portfolio. Deep neural networks are especially suited for this task as a huge amount of historical consumption data is available from smart meter recordings to be used for model training. Probabilistic layers further enable the estimation of the uncertainty of the consumption forecasts. Here, we propose a method to calculate hourly day-ahead energy consumption forecasts which include an estimation of the aleatoric uncertainty. To consider the statistical properties of energy consumption values, the aleatoric uncertainty is modeled using lognormal distributions whose parameters are calculated by deep neural networks. As a result, predictions of the hourly day-ahead energy consumption of single customers are represented by random variables drawn from lognormal distributions obtained as output from the neural network. We further demonstrate, how these random variables corresponding to single customers can be aggregated to probabilistic forecasts of customer portfolios of arbitrary composition.
\end{abstract}


\section{Introduction}
Predictions of energy consumption are crucial for energy retailers to minimize deviations from energy acquired in the day-ahead market and the actual consumption of their customers. The increasing spread of smart meters means that retailers have access to hourly consumption values of all their contracted customers in real time \cite{zheng2013}. Using machine learning algorithms, these hourly values can be used to calculate predictions for the future energy consumption of the customers \cite{amasyali2018, wang2019a}. However, many aspects of human behavior are not predictable only knowing the past consumption. There is always a \textit{known unknown} component in the future consumption which is called \textit{aleatoric uncertainty} \cite{hullermeier2021, gneiting2014}. Simply predicting the future consumption as one single value ignores this aleatoric uncertainty and will therefore always be limited in terms of accuracy. Hence, it is reasonable to include the aleatoric uncertainty into the prediction \cite{hong2016}. An approach to estimate the aleatoric uncertainty is to use prediction models whose outputs are distributions instead of single values. Here, we employ neural networks with probabilistic nodes in the output layer to calculate predictions of the customers energy consumption.
\\

For the predictions to be useful for retailers, it is necessary to provide not only forecasts for single customers, but also for customer portfolios of various composition. The prediction model should further be flexible enough, to allow for continuously changing portfolio members as customers change their retailers frequently. An option to fulfill these requirements is to calculate predictions for each customer separately and to afterwards aggregate all predictions to a forecast for the respective customer portfolio. While the aggregation of single predicted values can be conducted by simply calculating their sum, the aggregation of distributions is more complex. In essence, it depends on the type of distribution used inside the model. For the case that the consumptions are modeled using normal distributions, the sum of normally distributed random variables is again a normal distribution and there is a closed form for calculating the mean and standard deviation of the resulting aggregated distribution. Energy consumptions however do not follow a normal distribution. More adequate are skewed distributions like lognormal, Gamma- or Weibull distributions \cite{kuusela2015, carpaneto2008, munkhammar2014}. For these distributions, there is no closed form to calculate their aggregation. A possibility to calculate the aggregated probability density function is to calculate the convolution of the probability density functions of the distribution to be aggregated. While calculating such a convolution for all customers sequentially is computationally expensive, the parallel convolution of all probability density functions by calculating their product in frequency space is numerically unstable. Here, we therefore conduct the aggregation by sampling from all distributions to be aggregated, calculating the sum of the obtained random values and afterwards extracting the percentile values from the resulting summed random values. To enhance the reproducibility of this work, we publish the source code \cite{our_github} as well as the dataset \cite{our_dataset} used for the presented investigations along with this study. 
\begin{figure*}[ht]
\includegraphics[width=1.0\linewidth]{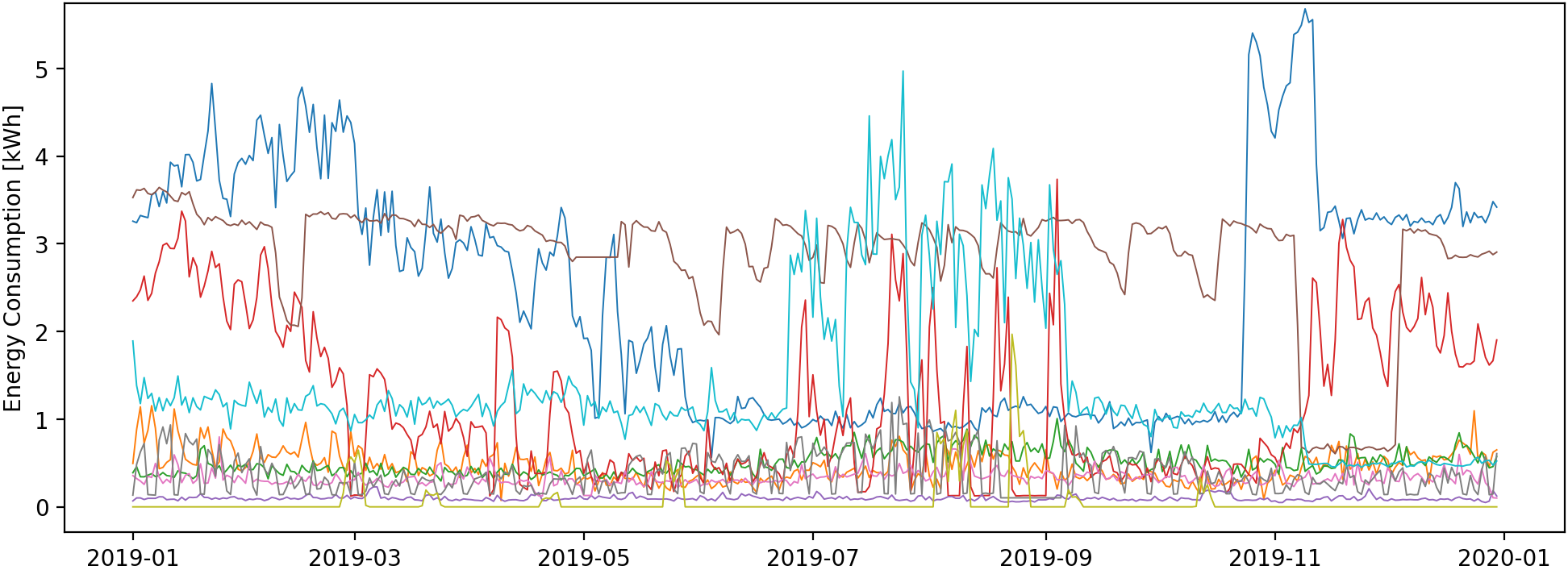}
\caption{Energy consumption curves for 10 example household customers}
\label{fig_consumption_examples}
\end{figure*}
\section{Methods}
\label{section_methods}
\subsection{Related literature}
For the prediction of energy consumptions, many works study the application of machine learning techniques to achieve accurate predictions of the future energy consumption \cite{robinson2017, truong2021, deng2018}. Given enough data for model training, deep learning can lead to even more accurate prediction models \cite{amasyali2018, li2017, mocanu2016}. Recent approaches employed hybrid recurrent-convolutional models \cite{sajjad2020}, recurrent neural networks \cite{shi2018} or residual architectures \cite{chen2019}. Methods which consider the remaining uncertainty have been developed for energy prices \cite{nowotarski2018}, generation as well as consumption \cite{meer2018}. For the case of energy consumption forecasting, the usage of additive quantile regression \cite{taieb2016}, Bayesian deep learning \cite{sun2020} as well as pinball loss guided long short-term memory networks \cite{wang2019b} was reported. Quantile regression averaging was used in \cite{wang2019c} to combine multiple probabilistic energy consumption predictions.
\subsection{The Dataset}
\begin{figure*}[ht]
\includegraphics[width=1.0\linewidth]{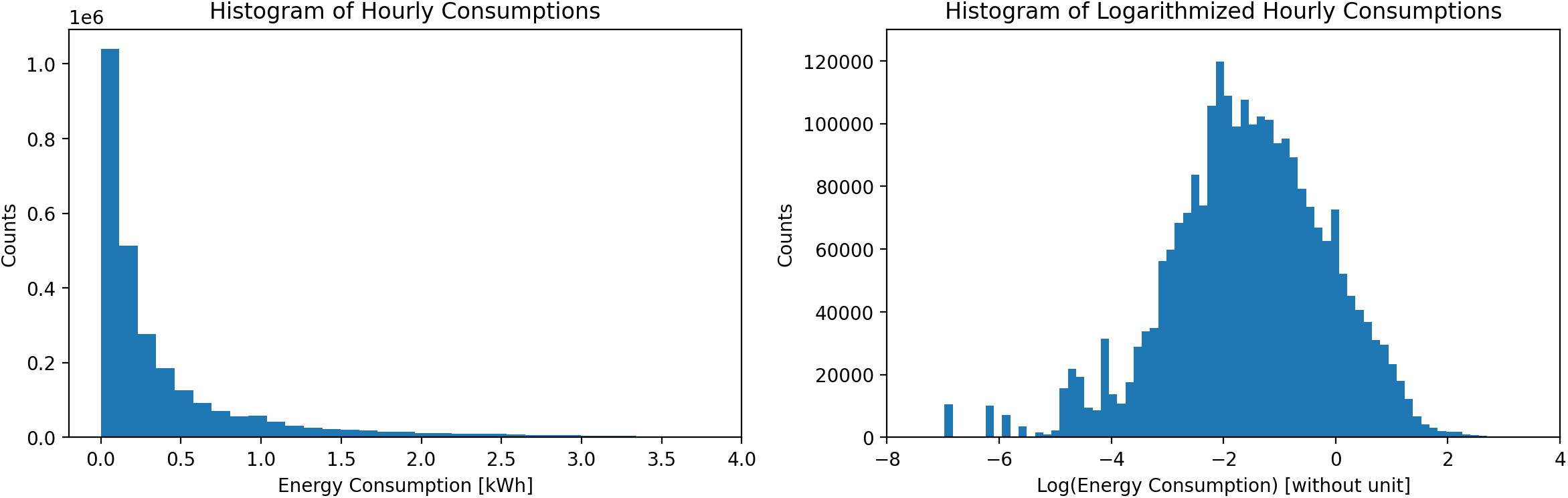}
\caption{Left: Histogram of the hourly energy consumptions of the household customers of the used dataset. Right: Logarithmized hourly energy consumptions of the household customers of the used dataset.}
\label{fig_consumption_distribution}
\end{figure*}
Basis for this work was a real-world dataset with energy consumption values of 499 anonymized customers located in Spain for the year 2019 in hourly resolution. In addition, temperature values corresponding to the zip code-exact location of each customer was part of the dataset. They had the same time range and resolution as the consumption values. The 499 customers comprise households as well as companies from various industrial sectors. Since consumption amounts for industrial customers lie in ranges magnitudes different from those of household customers, it was decided to focus on the 314 household customers inside the dataset. The consumption curves for 10 example customers are shown in Figure \ref{fig_consumption_examples}. It is apparent that the consumption curves from different customers vary in many kinds. Some show repeating weekly patterns while others stay at a constant consumption level for a long time or behave in more stochastic way. First investigations of the dataset showed that the household consumption values follow a lognormal distribution, as demonstrated in Figure \ref{fig_consumption_distribution}. In its left part the distinct consumption values are plotted as a histogram which showes a skewed distribution. For the case that the logarithm of the energy consumption values is plotted (right part of Figure \ref{fig_consumption_distribution}), the resulting values roughly follow a normal distribution. Still, it needs to be considered that the lognormal distribution only contains values greater than zero, whereas energy consumption values can actually be exactly zero.
\subsection{Preprocessing}
At first, the time changes in spring and autumn between Central European Time (CET) and Central European Summer Time (CEST) and vice versa necessitated adaptations to have 24 consumption values also at these days. For the spring time change, where the value at 2 am is missing, the average of the values at 1 am and 3 am was added at 2 am. For the autumn time change, where two values at 2 am exist, the average of both was taken as value for a single entry at 2 am.\\
In the next step, the dataset was enriched with information that might be useful for the prediction in addition to the already included consumption and temperature values. To allow the forecast algorithm to learn possible apparent weekly seasonalities, a one-hot-encoded day category was added. The used categories were “Monday”, “Tuesday-Thursday”, “Friday”, “Saturday” and “Sunday or Holiday”. For the category “Sunday or Holiday” it was necessary to extract the dates of Spanish public holidays, which differ, however, between the individual provinces. Since due to the anonymization no information about the location of the customers was available, dates of public holidays were assumed to be those of the Spanish province Madrid.\\
In addition, the number of the month (1-12) as well as the number of the day in the month (1-31) was added to enable the forecast algorithm to learn yearly seasonalities. Also, a temperature forecast column was added. Since no historical temperature forecast values were available, the temperature column was shifted by one day and used as temperature forecast. In a real application, actual temperature forecasts can then be used for predictions, since they are much easier to obtain compared to historical forecast values. Since 1-day temperature forecasts are quite exact nowadays, it was assumed that no major overestimation of the model’s accuracy is introduced by this approximation of the temperature forecast. Further, daily means of consumption and temperature forecasts were calculated.\\
The obtained dataset was splitted with respect of the 
individual customers into a train, test and validation dataset. The train data thereby contains randomly selected 252 customers (80 \% of 314 household customers), whereas both the test and the validation dataset each contain disjunct data of 31 customers (10 \%). The last preprocessing step comprised the scaling of the data. Since the consumption was later modeled using lognormal distributions, the scaled consumption values had to be larger than zero as well. To achieve a scaling robust to outliers, which maintains the positivity of the consumption values, the interquantile range between the quantiles q$_0$ and q$_{75}$ of the consumption values was extracted and the consumption data was divided by this interquantile range. A small increment of $\varepsilon=10^{-5}$ was added to the consumption values to avoid values which equal zero. This is necessary as the lognormal distribution used for modeling is only defined for positive real numbers. All the other columns had not the requirement to only have positive values which is why each of these features were centered to have zero mean and their scale was altered by dividing them by the interquantile range between the quantiles q$_{25}$ and q$_{75}$. All the parameters for the scaling and centering were calculated using the training dataset. The test and validation datasets were centered and scaled using these parameters from the training dataset.
\subsection{Consumption Forecast Model}
Since for this work a reasonable amount of training data was available, it was decided to use deep learning models to predict the day-ahead energy consumption. In regression tasks, as with the energy consumption prediction conducted here, neural networks are usually trained by minimizing the mean squared error (MSE) between the predicted value $y$ and the actual one $\hat{y}$. However, unless the MSE can be minimized to equal zero, a certain deviation between predicted and actual values will remain. To take this aleatoric uncertainty into account, in this work the prediction output is not a single value but a distribution. Tensorflow Probability \cite{dillon2017} was used for the implementation of the probabilistic layers. It was chosen to use a lognormal distribution according to the evaluation in Figure \ref{fig_consumption_distribution}. It has the probability density function
\begin{equation}
f(z)=\frac{1}{\sqrt{2\pi}\sigma z}\exp \left( - \frac{\left( \ln z -\mu \right)^2}{2\sigma ^2} \right)
\end{equation}
with $\mu$ and $\sigma$ the mean and standard deviation of the underlying normal distribution. Neural networks with probabilistic outputs can be optimized by minimizing the negative log-likelihood $L_{\mathrm{NLL}}$ of the target value y and the predicted distribution under given input x:
\begin{equation}
L_{\mathrm{NLL}} = - \log P(y|x).
\end{equation}
The neural network thereby provides the parameters $\mu$ and $\sigma$ of the respective output distribution per input x.
\\
\subsubsection{Branch A}
\begin{figure}[t]
\includegraphics[width=1.0\linewidth]{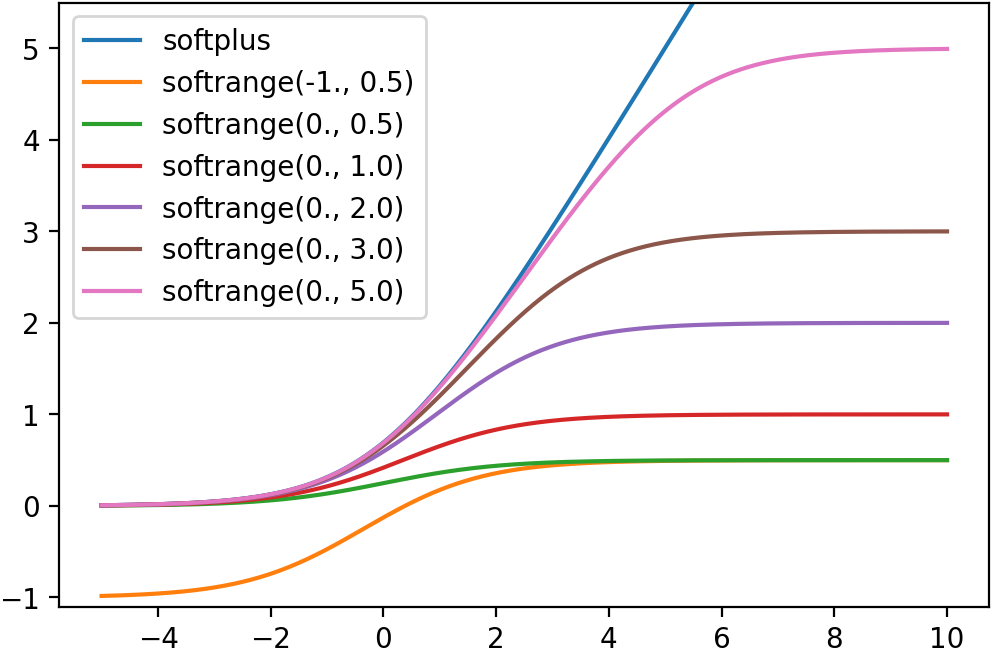}
\caption{Examples of differently parameterized softrange functions in comparison with the softplus function.}
\label{fig_softrange}
\end{figure}

~\\
\label{section_branch_a}
\begin{figure*}[ht]
\includegraphics[width=1.0\linewidth]{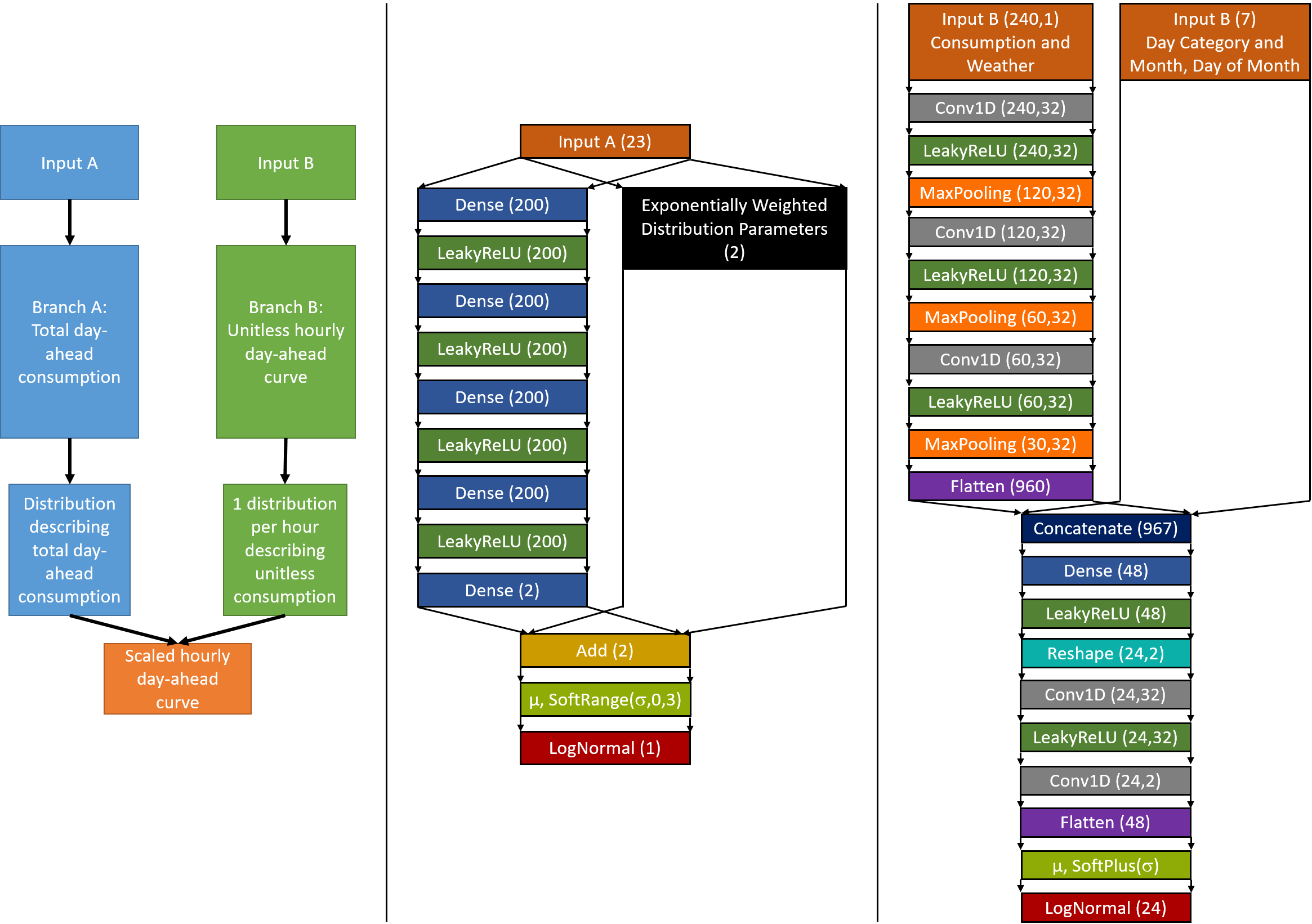}
\caption{Left: Sketch of the overall architecture of the neural network used to calculate hourly energy consumption predictions with an estimation of the aleatoric uncertainty. Center: Layer-by-layer implementation of branch A. Right: Layer-by-layer implementation of branch B.}
\label{fig_architecture}
\end{figure*}
For providing day-ahead forecasts in hourly resolution, we used a two-branch neural network architecture as depicted in the left part of Figure \ref{fig_architecture}. A branch A is thereby trained to predict the total amount of energy consumed during one day. It receives the daily mean values of consumption and weather forecast of the 2 weeks / 14 days prior to the day where the forecast shall be calculated. Using these inputs, it is assumed that the neural network is able to evaluate up to which extent the consumption of a specific customer correlates with the temperature. To be able to consider seasonal components, the neural network gets the one-hot encoded day category and the month number as well as the day of month number of the day for which the prediction is calculated. The branch A neural network itself comprises two branches as shown in the central part of Figure \ref{fig_architecture}. One branch is a multilayer perceptron (MLP) with 4 hidden layers, each with 200 nodes and LeakyReLU activation function, which receives the described variables as input. The second part is an exponentially weighted estimation of the lognormal distribution parameters of the past 2 weeks. Given a set of $n$ observed random values $X_i$, the lognormal distribution parameters can be estimated empirically as
\begin{equation}
\mu_{\mathrm{emp}}=\frac{1}{n}\sum^n_{i=1} \ln X_i
\label{eq_mu_emp}
\end{equation}
and
\begin{equation}
\sigma_{\mathrm{emp}}=\frac{1}{n-1}\sum^n_{i=1} \left( \ln X_i -\mu_{\mathrm{emp}} \right)^2.
\label{eq_sigma_emp}
\end{equation}
With the exponential weighting, which gives more impact to the most recent values, \eqref{eq_mu_emp} and \eqref{eq_sigma_emp} become:
\begin{equation}
\mu_{\mathrm{emp,exp}}=\frac{1}{\sum^n_{i=1} c_{\mu,i}}\sum^n_{i=1} c_{\mu,i}\cdot \ln X_i
\end{equation}
and
\begin{equation}
\sigma_{\mathrm{emp,exp}}=\frac{1}{\sum^n_{i=1} c_{\sigma,i} \cdot \left( 1- \frac{1}{n} \right)}\sum^n_{i=1} c_{\sigma,i}\cdot \left( \ln X_i -\mu_{\mathrm{emp}} \right)^2.
\end{equation}
with the exponentially decaying weights
\begin{equation}
c_{\mu,i} = c_\mu(t_i)=e^{-\lambda_\mu \cdot t_i}
\label{eq_c_mu_i}
\end{equation}
and
\begin{equation}
c_{\sigma,i} = c_\sigma(t_i)=e^{-\lambda_\sigma \cdot t_i}.
\label{eq_c_sigma_i}
\end{equation}
$t_i \in [ 0,1,...,13]$ thereby describes the difference in days between the dates of the most recent observation $X_0$ and the observation $X_i$ at time $t_i$. The parameters $\lambda_\mu$ and $\lambda_\sigma$ are parameters to be optimized by the neural network. This means, that the neural network is able to optimize how much influence it provides to the most recently observed values in comparison to the values which were observed several days before. The results of the MLP and the exponentially weighted distribution parameter estimation are afterwards added. Due to this addition, the MLP only needs to learn the residual part of the already estimated distribution parameters which led to results with higher accuracy in our experiments.\\
The predicted scale parameter $\sigma$ of the output distribution needs to be constrained into a range of allowed values before passing it into the lognormal distribution layer. Firstly, it needs to be larger than zero due to the properties of the lognormal distribution. And secondly, it should be constrained to not take on a value which causes extremely broad distributions. For some data points, especially those which are of higher difficulty for the neural network to predict, it is easier for the neural network to just largely increase the scale parameter $\sigma$ of the output distribution rather than learning a better fit for the location parameter $\mu$. However, this behavior complicates the aggregation of the prediction for multiple customers later on. The scale parameter was therefore constrained to have maximum value of $\sigma_{\mathrm{max}}=3$. Simply replacing values outside the targeted interval $[0,\sigma_{\mathrm{max}}]$ with these minimum or maximum values would lead to gradients which equal zero during error backpropagation of samples where this replacement is necessary. To maintain gradients for all possible values of $\sigma$ which are output by the neural network, the constraint needs to be conducted using a differentiable function. Here we propose the \textit{softrange} function $f_{\mathrm{sr}}$, which can be calculated as follows:
\begin{IEEEeqnarray}{rCl}
f_{\mathrm{sr}}(x, y_{\mathrm{l}}, y_{\mathrm{u}})& = & y_{\mathrm{l}}+\frac{y_{\mathrm{u}}-y_{\mathrm{l}}}{f_{\mathrm{s+}}(y_{\mathrm{u}}-y_{\mathrm{l}})} \nonumber \\
&& \cdot f_{\mathrm{s+}}( -f_{\mathrm{s+}}(-x+y_{\mathrm{u}}) +y_{\mathrm{u}} -y_{\mathrm{l}})
\end{IEEEeqnarray}
Here, $y_{\mathrm{l}}$ and $y_{\mathrm{u}}$ are the aspired lower and upper boundaries and $f_{\mathrm{s+}}$ is the softplus function
\begin{equation}
f_{\mathrm{s+}}(x)=\ln (1+ e^x)
\end{equation}
which is a common activation function for neural networks. Some example plots of differently parameterized softrange functions in comparison with the softplus function are given in Figure \ref{fig_softrange}.
\\
\subsubsection{Branch B}~\\
The layer-by-layer architecture of branch B is depicted in the right part of Figure \ref{fig_architecture}. Here, the optimization task was to predict the unitless intraday consumption curve for the next day. Unitless means, that during training of the intraday curves, they were scaled to sum up to a value of 24 (on average a value of 1 per hour). This means that the prediction of the intraday curve is totally independent of the total amount of energy consumed at that day. Since the amount of energy consumed is predicted by branch A, branch B could completely focus on consumption patterns of each customer like a higher consumption during daytimes compared to the hours at night. As the prediction tasks are unitless here, the task for branch B is mostly a pattern recognition and pattern creation task. Convolutional neural networks are especially suited for these pattern-related tasks which is why branch B is composed of convolutional layers. It receives hourly consumption data of the seven days prior to the day to predict as well as hourly temperature forecast data of the three days prior to the day of prediction. These inputs are processed by three subsequent blocks each composed of a 1D convolution with a filter size of 5, a LeakyReLU activation and a MaxPooling with a pool size of 3 and a stride of 2. After those three blocks the tensor is flattened and concatenated with a tensor that contains the day category, the number of the month as well as the number of the day inside the month. The resulting tensor is then processed by a fully connected layer with 48 nodes and a LeakyReLU activation function. It can thereby combine the information, which was extracted by the convolutional layers through processing of the consumption and forecast data, and the information related to the day of prediction added in the concatenation before. After a necessary reshaping it is again passed through convolutional layers to create the pattern of hourly $\mu$ and $\sigma$ values for the final output distributions. Since the outputs of branch B are scaled later so that they sum up to the output value of branch A, it is not necessary here to introduce a maximum value for the scale parameters $\sigma$. Therefore, the $\sigma$ values are only passed through a softplus function to constrain them to positive values in a differentiable way. The resulting 24 values for $\mu$ and 24 values for $\sigma$ are then passed to a LogNormal layer to create the 24 output distributions out of them, one for each hour of the following day.\\
\subsubsection{Scaling the Output of Branch B Using the Output of Branch A}~\\
To get the hourly predictions of the day-ahead consumption, it is necessary to scale the 24 random variables describing the unitless consumption curve provided by branch B with the total day-ahead consumption predicted by Branch A. This means that the sum of the 24 random variables from branch B must equal the statistical properties of the random variable which is the output of branch A. Unlike the case for normally distributed random variables, there is no closed form for calculating the resulting distribution of the sum of multiple lognormally distributed random variables. We therefore approximated the statistical properties of the sum of all 24 random variables provided by branch B by an approach based on random sampling. At first, we sampled a large number $n_{\mathrm{samples}}=5000$ of random numbers from each of the 24 lognormal output distributions of branch B, which have the parameters $\mu_{\mathrm{B},k}$ and $\sigma_{\mathrm{B},k}$ with $k \in [ 0,1,...,24]$. We then calculated $n_{\mathrm{samples}}$ summed values out of these $24\cdot n_{\mathrm{samples}}$ random numbers, which is equivalent to sampling directly from an aggregated distribution of the 24 output distributions. As a next step, we extracted the mean $E_{\mathrm{agg,B}}$ and the median $\mu^*_{\mathrm{agg,B}}$ from the $n_{\mathrm{samples}}$ summed random numbers. The median thereby is a scalable parameter, which means that
\begin{equation}
\mu^*(a\cdot X)=a\cdot \mu^*(X)
\end{equation}
with an arbitrary constant $a$ and a random variable $X$. Together with the median of the output distribution from branch A, $\mu^*_{\mathrm{A}}$, we can calculate a scale parameter $a_{\mu}$, which represents the ratio between $\mu^*_{\mathrm{A}}$ and $\mu^*_{\mathrm{agg,B}}$:
\begin{equation}
a_{\mu}=\frac{\mu^*_{\mathrm{A}}}{\mu^*_{\mathrm{agg,B}}}.
\label{eq_a_mu}
\end{equation}
By using \eqref{eq_a_mu}, we can scale the unitless hourly median values so, that their aggregation equals $\mu^*_{\mathrm{A}}$:
\begin{equation}
\mu^*_{\mathrm{scaled},k}=a_{\mu}\cdot \mu^*_{\mathrm{B},k} .
\end{equation}
This enables us to calculate the parameters $\mu_{\mathrm{scaled},k}$ of the 24 distributions per forecast which describe the predicted hourly consumption:
\begin{equation}
\mu_{\mathrm{scaled},k} = \ln (\mu^*_{\mathrm{scaled},k}) = \ln (a_{\mu} \cdot \mu^*_{\mathrm{B},k} ).
\end{equation}
To calculate the parameters $\sigma_{\mathrm{scaled},k}$ of the scaled intraday forecast distributions, we assume that $n_{\mathrm{samples}}$ is large enough that $E_{\mathrm{agg,B}}$ is a good approximation for the expected value of the aggregated distribution of the unitless intraday consumption curve calculated by branch B. Like the median, the expected value of a distribution is a scalable parameter as well. Like for the median, we can calculate the 24 expectation values $E_{\mathrm{scaled},k}$ for the scaled hourly distributions:
\begin{equation}
E_{\mathrm{scaled},k} = a_E \cdot E_{\mathrm{B},k} = \frac{E_{\mathrm{A}}}{E_{\mathrm{agg,B}}} \cdot E_{\mathrm{B},k}.
\end{equation}
Here, $E_{\mathrm{A}}$ is the expected value of the distribution predicted by branch A and $E_{\mathrm{B},k}$ are the 24 expectation values from the distributions predicted by branch B. Since the expected value $E$ of a lognormal distribution is given by
\begin{equation}
E = e^{\mu + \sigma^2}
\end{equation}
we can calculate the scale parameters $\sigma_{\mathrm{scaled},k}$ of the distributions for the hourly consumption forecast:
\begin{equation}
\sigma_{\mathrm{scaled},k} = \sqrt{2\cdot \left( \ln (E_{\mathrm{scaled},k}) - \ln (\mu_{\mathrm{scaled},k}) \right)}.
\end{equation}
With having $\mu_{\mathrm{scaled},k}$ and $\sigma_{\mathrm{scaled},k}$ calculated, one can easily extract the upper and lower boundaries $\hat{y}_{\mathrm{u},k}$ and $\hat{y}_{\mathrm{l},k}$ of the predictions for the hourly consumption:
\begin{equation}
\hat{y}_{\mathrm{u},k} = e^{\mu_{\mathrm{scaled},k}} \cdot e^{\sigma_{\mathrm{scaled},k}}
\end{equation}
and
\begin{equation}
\hat{y}_{\mathrm{l},k} = e^{\mu_{\mathrm{scaled},k}} / e^{\sigma_{\mathrm{scaled},k}}.
\end{equation}
Given a model, which describes the aleatoric uncertainty perfectly, the predicted intervals $[\hat{y}_{\mathrm{l},k}, \hat{y}_{\mathrm{u},k}]$ would contain roughly 68 \% of the actual consumption values. At the end, the increment $\varepsilon$, which was added in the preprocessing part, was subtracted from the extracted median values as well as the upper and lower boundaries.\\
\subsubsection{Aggregation of Predictions for Multiple Customers}~\\
To aggregate consumptions of multiple customers, either on a daily or an hourly basis, again a sampling procedure is utilized. For all distributions to be aggregated, again $n_{\mathrm{samples}}=5000$ random numbers are sampled, and summed up along the axis where aggregation shall be conducted. From the $n_{\mathrm{samples}}$ aggregated random numbers, we extracted the median as well as the upper and lower boundaries of the resulting aggregated distributions by extracting the corresponding quantiles q$_{50}$, q$_{84.135}$ and q$_{15.865}$. As with the scaling part, the increment $\varepsilon$ needs to be subtracted for the final values.
\section{Results}
\begin{figure*}[ht]
\includegraphics[width=1.0\linewidth]{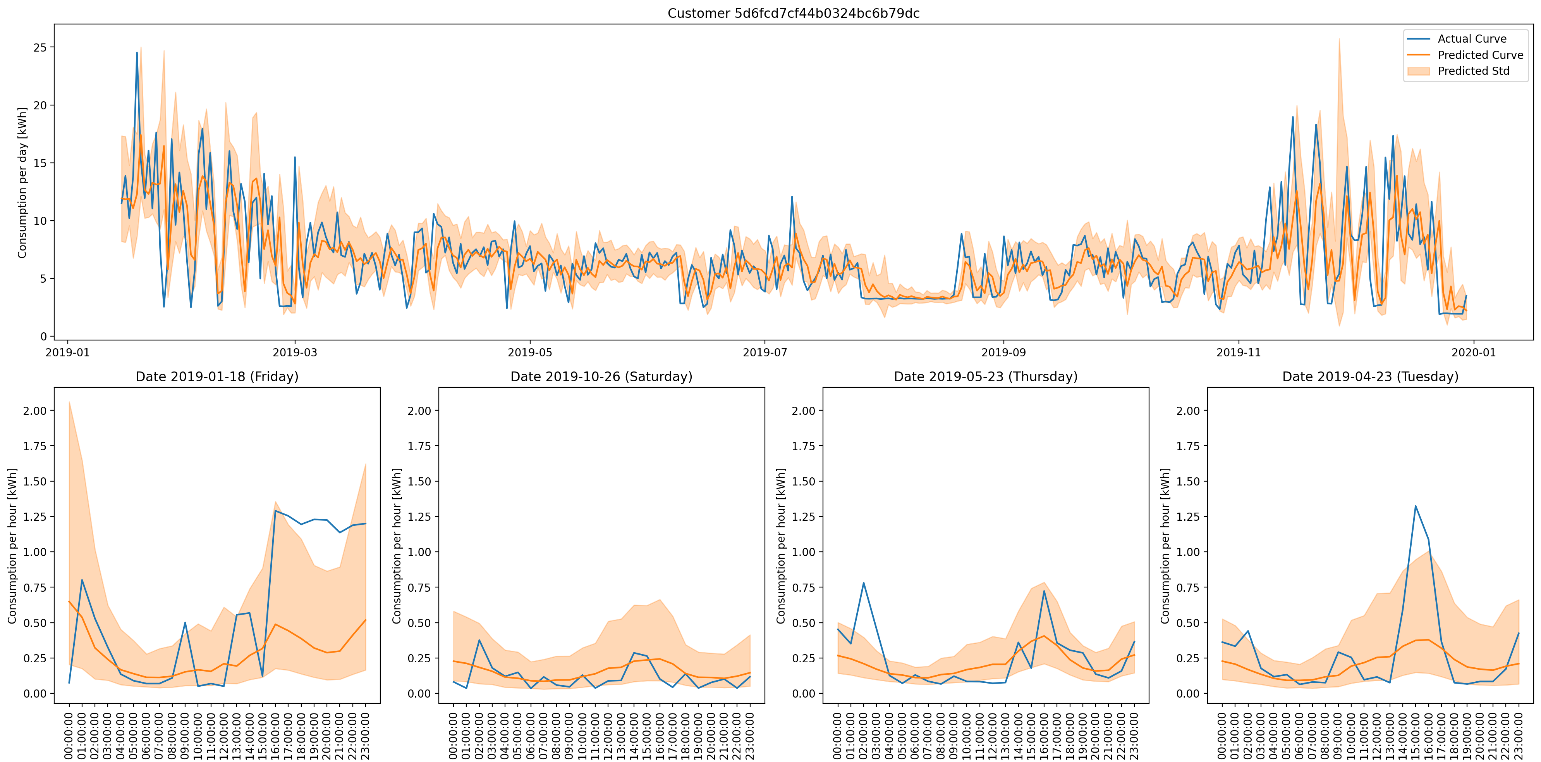}
\caption{Results for an example customer from the validation dataset in daily resolution (upper part) and in hourly resolution for randomly selected example days.}
\label{fig_results_example_customer}
\end{figure*}

\begin{figure*}[ht]
\includegraphics[width=1.0\linewidth]{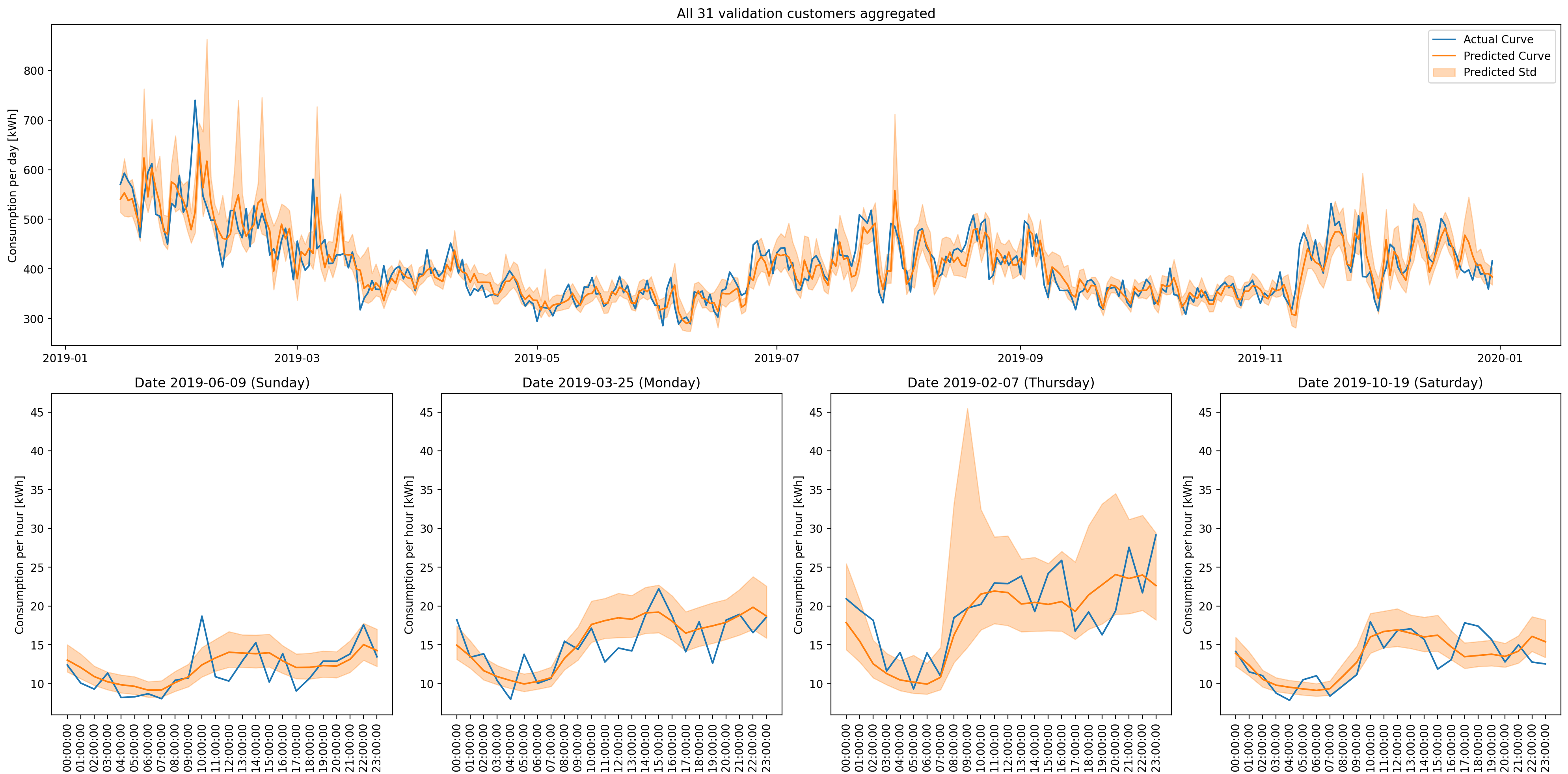}
\caption{Results for the aggregated portfolio of 31 validation customers in daily resolution (upper part) and in hourly resolution for randomly selected example days.}
\label{fig_results_aggregated}
\end{figure*}

An example of the prediction results for a customer from the validation dataset, which was not used for training of the model, is given in Figure \ref{fig_results_example_customer}. In its upper part, the actual values, the predicted values as well as the predicted intervals in which roughly 68 \% of all actual values should lie, are plotted. It is apparent, that the proposed algorithm is able to fit to periods with a strong weekly periodicity (for example in February 2019) as well as to periods with a constant consumption (August 2019). Of course, for cases with rapidly changing behavior it needs 1-2 days to adapt (December 2020). The uncertainty intervals clearly adapt to the changing uncertainty about the customers energy consumption. While the uncertainty is large in March 2019, it decreases from end of July to mid of August 2019, when the customer had a constant consumption over a long time period. Looking at the hourly predictions, it is suspicious that the uncertainty intervals are much larger due to the higher uncertainty in the hourly consumption. The prediction model is able to differentiate between hours of a day with higher and lower uncertainty of their energy consumption. Due to the increased uncertainty, the matching between the predicted and the actual energy consumption is worse compared to the daily values.\\
The aggregation of all 31 validation customers is shown in Figure \ref{fig_results_aggregated}. Both for the daily as well as the hourly resolution the area between the uncertainty boundaries is much lower in relation to the predicted consumption values compared to the case of a single customer. This is an intended behavior, since unforeseeable changes in single customer’s energy consumption average out for the aggregation of a large number of customers.\\
\begin{table*}[!b]
\caption{Evaluation Metrics}
\begin{tabularx}{\textwidth}{@{}|Y|Y|Y|Y|Y|@{}}
\hline
\textbf{Metric}&\textbf{Median relative error (training, 252 customers)}&\textbf{Median relative error (validation, 31 customers)}&\textbf{Percentage of actual values inside uncertainty interval (training, 252 customers)}&\textbf{Percentage of actual values inside uncertainty interval (validation, 31 customers)}\\
\hline
Single customers, hourly resolution & 30.9 \% & 28.1 \% & 77.4 \% & 77.7 \% \\ \hline
Single customers, daily resolution & 11.7 \% & 11.0 \% & 76.0 \% & 74.5 \% \\ \hline
Customer portfolio, hourly resolution & 5.4 \% & 11.2 \% & 69.0 \% & 64.1 \% \\ \hline
Customer portfolio, daily resolution & 2.5 \% & 4.3 \% & 62.0 \% & 71.4 \% \\ \hline
\end{tabularx}
\label{table_results_metrics}
\end{table*}
As discussed in Section \ref{section_branch_a}, the parameters $\lambda_{\mu}$ and $\lambda_{\sigma}$ determine the decay rate of the exponentially weighted distribution parameter estimation in branch A. As they are optimized by the branch A neural network, it is possible to extract their value after the training phase and to evaluate, whether the neural network focused only on the most recent values or whether it utilized the whole set of historical consumption values given as input. The decay parameter $\lambda_{\mu}$ was optimized to a value of 1.09. According to \eqref{eq_c_mu_i}, this means that the consumption values directly prior to the day of prediction were weighted more than 1 million times stronger than the consumption values 2 weeks prior to the day of prediction. In other words, the neural network mainly focused on the most recent consumption values for having a first estimation of the location parameter $\mu$. The decay rate $\lambda_{\sigma}$, which determines the decay related to the exponentially weighted estimation of the scale parameter, was optimized to a value of 0.09. According to \eqref{eq_c_sigma_i}, this means that the consumption values directly prior to the day of prediction were weighted about 3 times stronger than the consumption values 2 weeks prior to the day of prediction. It shows that for the estimation of the scale parameter, the neural network utilized the full available information about the consumption history. It is also a sign, that the estimation of uncertainty intervals could benefit from an extension of the consumption time range, which is fed to branch A.\\
In addition to the visual evaluation, the results in the different aggregation levels were also evaluated using the metrics given in Table \ref{table_results_metrics}. To evaluate the accuracy of the energy consumption forecasts without considering the also predicted uncertainty, we used the median relative error (MdRE) of the actual values and the respective medians of the predicted distributions. The reason for not using the mean relative error (MRE) of the actual values and the respective medians of the predicted distributions is that some actual consumption values equal zero which leads to numerically unstable results of the MRE. To evaluate the correctness of the predicted uncertainty intervals, we calculate the percentage of actual values, which lie between the predicted upper and lower border. A perfect result here would be a value of roughly 68 \%. We evaluated both metrics for the training as well as for the validation dataset. For the MdRE, we see a clearly decreasing error with increasing level of aggregation. For the case of single customers, the error values are nearly the same on the training and the test dataset. This changes for the case of the customer portfolios, where the predictions of all customers of the respective datasets are aggregated. Here we have a lower error on the training dataset. However, this does not necessarily mean that the model overfitted on the training data, because due to the larger amount of customers in the training dataset, the level of aggregation is higher in comparison with the aggregated validation dataset. For comparison, the strategy of taking the consumption value of the day prior to the day of prediction as forecast would lead to MdREs of 2.75 \% and 5.58 \% for the training and validation dataset, respectively, when the aggregation level of daily resolution and customer portfolios is used.\\
The metric-based evaluation of the uncertainty intervals proves, that the proposed method to predict uncertainty estimations is stable with regard to the aggregation of a high number of independent instances. Still, there is space for improvement since the targeted value of 68 \% is missed up to 10\%-points. As with the MdRE, no overfitting is visible as the results for both the training and the validation dataset are similar.
\section{Discussion and Conclusion}
Using the proposed method, it is possible to obtain predictions for energy consumptions recorded by smart meters for the following day as well as estimations of the aleatoric uncertainty of these predictions. Since one forecast per smart meter instance is calculated and also the aggregation of a high number of customers is accurate using the proposed method, it is possible to generate forecasts for arbitrary customer portfolios by aggregating the single customer predictions in the described way.\\
A limitation of this method is that the scaling and aggregation procedure is based on random sampling. This means that the calculated median and uncertainty interval results will never be as accurate as an analytical solution would be. However, since random sampling is a computationally relative cheap operation, increasing the number of samples per aggregation to a very huge number is always possible. In addition to an improvement regarding the metrics discussed here, further work should also consider to include the \textit{epistemic uncertainty}, which describes the uncertainty of the weights of the model as this type of uncertainty remained unconsidered in this work. The usage of further data sources could also lead to improved results. For example, on can expect that a high level of precipitation or air humidity leads to an increase in the duration, people stay inside their respective homes. This, in turn, could lead to a higher energy consumption at these times.

\end{document}